\documentclass[prl,twocolumn,amsmath,amsfonts,superscriptaddress,floatfix]{revtex4-1}

\usepackage{amssymb}
\usepackage[mathscr]{eucal}
\usepackage[dvips]{graphicx}
\usepackage{color}
\usepackage{bbold}

\newcommand{\ket}[1]{| \, #1 \, \rangle}
\newcommand{\bra}[1]{\langle \, #1 \, |}

\newcommand{\scal}[2]{\bra{#1} \, #2 \, \rangle}
\newcommand{\expect}[1]{\langle #1 \rangle}

\begin{document}

\title{\fontsize{12pt}{12pt} \selectfont Numerical Contraction of the Tensor Network generated by the \mbox{Algebraic Bethe Ansatz}}

\author{V. Murg}
\affiliation{\mbox{Vienna Center for Quantum Science and Technology, Faculty of Physics, University of Vienna, Vienna, Austria}}

\author{V. E. Korepin}
\affiliation{C. N. Yang Institute for Theoretical Physics,  State University of New York at Stony Brook, NY 11794-3840, USA}

\author{F. Verstraete}
\affiliation{\mbox{Vienna Center for Quantum Science and Technology, Faculty of Physics, University of Vienna, Vienna, Austria}}

\date{\today}

\begin{abstract}
The algebraic Bethe Ansatz is a prosperous and well-established method for solving one--dimensional 
quantum models exactly. 
The solution of the complex eigenvalue problem is thereby reduced to the solution of
a set of algebraic equations.
Whereas the spectrum is usually obtained directly, the eigenstates are available only in terms of
complex mathematical expressions. This makes it very hard in general to extract properties from the states,
like, for example, correlation functions.
In our work, we apply the tools of Tensor Network States to describe the eigenstates approximately as
Matrix Product States. From the Matrix Product State expression, we then obtain observables like 
correlation functions directly.
\end{abstract}


\maketitle


The Coordinate Bethe Ansatz~\cite{bethe31}, as originally invented by Bethe to give an exact solution to the one-dimensional
antiferromagnetic Heisenberg model, reduces the complex problem of diagonalizing the Hamiltonian to finding the solutions of a set of algebraic equations.
Once solutions to these algebraic equations are found -- numerical approaches to find them efficiently exist in many cases -- 
the eigenvalues are known exactly. However, the eigenstates are available only as a complex mathematical expressions.
This makes it insuperable to get interesting properties out of the states -- like their entanglement characteristics or
their correlations.

A complementary approach is the Algebraic Bethe Ansatz~\cite{korepin93} (also known as ``inverse scattering method'').
In this approach, the scattering matrix ($R$-tensor) is in focus. Based on this matrix, 
the Hamiltonian is derived and eigenstates are constructed.
The ``inverse'' problem consists in finding the scattering matrix that represents favored Hamiltonian.
As in case of the Coordinate Bethe Ansatz, the eigenvalue problem is reduced to solving a set
of algebraic equations. From the solutions, the eigenvalues are obtained directly.
The eigenstates are available only in terms of a non-contractable tensor network.
This makes exact calculations of expectation values in general unfeasable.
However, because of their inherent structure, it has been proven to be possible to calculate the norm
and the scalar product between Bethe states exactly~\cite{korepin93,slavnov07}.

\begin{figure}[t]
    \begin{center}
        \includegraphics[width=0.48\textwidth]{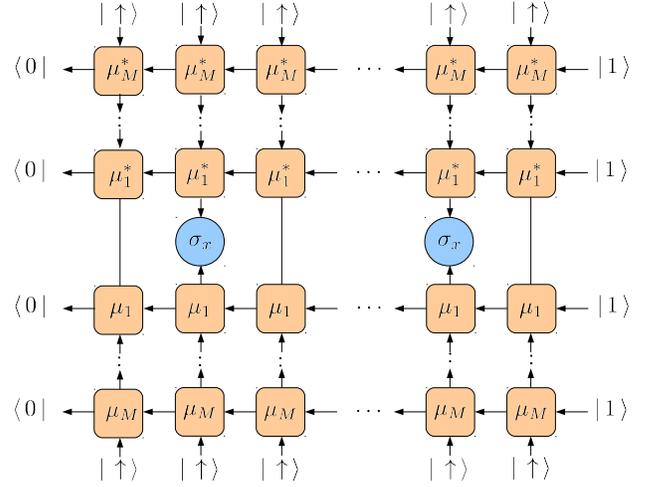}
     \end{center}
    \caption{
	Tensor network describing correlations $\expect{\sigma_x^{(i)} \sigma_x^{(j)}}$ 
	with respect to a Bethe-eigenstate.
       }
    \label{fig:betheansatzexp}
\end{figure}

In this paper, we take advantage of the fact that the Bethe-eigenstates have the form of a tensor-network~\cite{murg12,katsura10}.
The calculation of correlation functions then requires the contraction
of a tensor network such as the one depicted in Fig.~\ref{fig:betheansatzexp}.
We contract the tensor network approximately using a similar method as for time evolution 
in the Density Matrix Renormalization Group (DMRG)~\cite{white92,white92b,verstraeteciracmurg08,murgverstraete05}.
Finally, we end up with a Matrix Product State (MPS)~\cite{affleck87,affleck88,verstraetecirac05,perezgarcia06,verstraeteciracmurg08,singh10}
from which we can extract expectation values like correlation
functions directly. We show for the case of the antiferromagnetic Heisenberg model and the XXZ model with both periodic
and open boundary conditions that correlations can be obtained for~$50$ sites with good precision.


\begin{figure}[t]
    \begin{center}
        \includegraphics[width=0.48\textwidth]{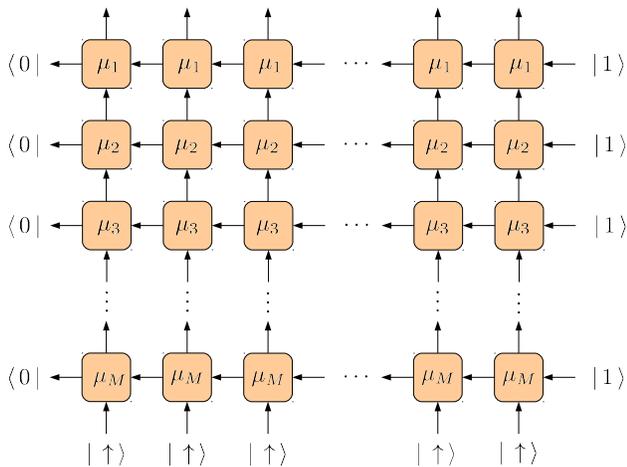}
     \end{center}
    \caption{
	Tensor network constituting the Bethe eigenstate of the
	Heisenberg model or XXZ model with periodic boundary conditions.
       }
    \label{fig:betheansatz}
\end{figure}

The Bethe-eigenstates 
for the antiferromagnetic Heisenberg model and the XXZ model with periodic boundary conditions
are obtained as
products of operators $B(\mu_j)$ applied on a certain vacuum state $\ket{vac}$, i.e.
\begin{equation} \label{eqn:betheansatz}
\ket{\Psi(\mu_1,\ldots,\mu_M)} = B(\mu_1) \cdots B(\mu_M) \ket{vac}.
\end{equation}
The parameters $\{\mu_j\}$ are thereby solutions of Bethe equations
and the $B(\mu_j)$'s play the role of creation operators of down-spins.
The vacuum corresponds to the state with all spins up.
$B(\mu)$ has the form of the Matrix Product Operator (MPO)~\cite{murg12}
\begin{displaymath}
B(\mu)
=
\sum_{\begin{smallmatrix}k_1 \cdots k_N\\l_1 \cdots l_N\end{smallmatrix} }
\bra{0}
\mathcal{L}^{k_1}_{l_1}(\mu)
\cdots
\mathcal{L}^{k_N}_{l_N}(\mu)
\ket{1}
o^{k_1}_{l_1} \otimes \cdots \otimes o^{k_N}_{l_N}
\end{displaymath}
with $k,l \in \{0,1\}$, $o^k_l = \ket{k}\bra{l}$ ($0 \equiv \uparrow$, $1 \equiv \downarrow$) 
and $\mathcal{L}^{k}_{l}(\mu)$ being $2 \times 2$
matrices dependent on the parameter~$\mu$.
The product of operators $B(\mu_1) \cdots B(\mu_M)$ can be read as
the contraction of the set of $4$-index tensors $[\mathcal{L}^{k}_{l}(\mu_j)]^{r}_{r'}$
with respect to a rectangular grid, as shown in Fig.~\ref{fig:betheansatz}.
Thereby, $r$, $r'$, $k$ and $l$ label the left, right, up and down-indices, respectively.
The non-zero entries of the tensor $[\mathcal{L}^{k}_{l}(\mu_j)]^{r}_{r'}$ are
\begin{displaymath}
\begin{array}{ccccc}
\small[\mathcal{L}^{0}_{0}(\mu)]^{0}_{0} & = & [\mathcal{L}^{1}_{1}(\mu)]^{1}_{1} & = & 1 \\
\small[\mathcal{L}^{0}_{1}(\mu)]^{1}_{0} & = & [\mathcal{L}^{1}_{0}(\mu)]^{0}_{1} & = & b(\mu) \\
\small[\mathcal{L}^{0}_{0}(\mu)]^{1}_{1} & = & [\mathcal{L}^{1}_{1}(\mu)]^{0}_{0} & = & c(\mu) \\
\end{array}
\end{displaymath}
with $b(\mu) = 1/(1+\mu)$, $c(\mu) = \mu/(1+\mu)$
for the Heisenberg model and
$b(\mu) = \sinh(2 i \eta)/\sinh(\mu+2 i \eta)$,
$c(\mu) = \sinh(\mu)/(\sinh(\mu+2 i \eta)$
for the XXZ-model ($\eta$ is related to the inhomogenity $\Delta$ in the XXZ model via $\Delta = \cos(2\eta)$).

The calculation of expectation values with respect to such a Bethe-eigenstate is
a considerably complex problem, because it requires the contraction of the tensor network depicted in Fig.~\ref{fig:betheansatzexp}.
This tensor network represents the correlation function
$\bra{\Psi(\mu_1,\ldots,\mu_M)} \sigma_x^{(i)} \sigma_x^{(j)} \ket{\Psi(\mu_1,\ldots,\mu_M)}$.
It is composed of the network for $\ket{\Psi(\mu_1,\ldots,\mu_M)}$ shown in Fig.~\ref{fig:betheansatz},
the vertically mirrored network with conjugated tensor entries representing $\bra{\Psi(\mu_1,\ldots,\mu_M)}$
and two operators $\sigma_x^{(i)}$ and $\sigma_x^{(j)}$ squeezed in-between at sites~$i$ and~$j$.
A tensor-network with such a structure also appears in connection with
the calculation of partition functions of two-dimensional classical systems and
one-dimensional quantum systems~\cite{murgverstraete05} and
the calculation of expectation values with respect to 
Projected Entangled Pair States (PEPS)~\cite{verstraetecirac04,murgverstraete07,murgverstraete08}.
The complexity to contract this network scales exponentially with the number of rows~$M$ or columns~$N$
(depending on the direction of contraction), which render practical calculations infeasible.

To circumvent this problem, we attempt to perform the contraction in an approximative numerical way:
the main idea is to consider the network in Fig.~\ref{fig:betheansatz} as the time-evolution of the state with all spins up
by the evolution operators $B(\mu_1),\ldots,B(\mu_M)$ to the final state $\ket{\Psi(\mu_1,\ldots,\mu_M)}$.
After each evolution step, the state remains a MPS, but the virtual dimension is increased by a factor of~$2$.
Thus, we approximate the MPS after each evolution step by a simpler MPS with smaller virtual dimension.
Of course, caution has to be used, because the operators $B(\mu_j)$ are not unitary and the
intermediate states of the evolution can be non-physical (i.e. they might have to be represented by MPS with high virtual dimension).
The Bethe-network, however, bears an additional structure which can be taken advantage of:
all evolution operators $B(\mu_j)$ commute, such that the operators $B(\mu_j)$ can be arbitrarily ordered.
This allows us to choose the optimal ordering for the evolution with intermediate states that are least entangled.
We will discuss this in detail the following.

The algorithm consists of~$M$ steps $m=1,\ldots,M$:
in the first step, the vacuum-state~$\ket{vac}$ is multiplied by the MPO~$B(\mu_1)$ 
to form the initial MPS~$\ket{\tilde{\Psi}_1}$.
In step~$m>1$, the product $\ket{\Psi_m} \equiv B(\mu_m) \ket{\tilde{\Psi}_{m-1}}$ is approximated
by the MPS $\ket{\tilde{\Psi}_m}$ that has maximal bond-dimension~$\tilde{D}$
and is closest to $\ket{\Psi_m}$.
In other words, we try to solve the minimization problem
\begin{equation} \label{eqn:optproblem}
K := \left\| \ket{\Psi_m} - \ket{\tilde{\Psi}_m} \right\|^2 \to \textrm{Min.}
\end{equation}
by optimizing over all matrices of the MPS~$\ket{\tilde{\Psi}_m}$.
This minimization problem also appears in the context of 
numerical calculation of expectation values with respect to PEPS~\cite{verstraetecirac04,murgverstraete07,murgverstraete08},
the calculation of partition functions~\cite{murgverstraete05} and (imaginary) time-evolution
of 1D quantum systems~\cite{verstraeteciracmurg08}.
We discuss it in detail in the supplementary part.
In this way, the MPS-approximation of the Bethe-state is obtained for $m=M$.
Thereby, $\{\mu_1,\ldots,\mu_M\}$ are the solutions of the Bethe-equations.
The error of the approximation is well-controlled in the sense that the expectation-value of
the energy can always be calculated with respect to the approximated MPS~$\ket{\tilde{\Psi}_M}$
and compared to the exact energy available from the Bethe-ansatz.

In order to make the algorithm more efficient, we take into account the
``creation operator''-property of the MPO~$B(\mu)$:
Since each MPO $B(\mu)$
creates one down-spin, the MPS~$\ket{\Psi_m}$ at step~$m$ contains exactly~$m$ down-spins.
Explicitly, the MPS reads~\cite{murg12}
\begin{displaymath}
\ket{\Psi_m} = \sum_{k_1 \cdots k_N} \bra{0} \bra{0} \mathcal{A}^{k_1} \cdots \mathcal{A}^{k_N} \ket{0} \ket{m} \ket{k_1,\ldots,k_N}
\end{displaymath}
with matrices $\mathcal{A}^k$ being block-diagonal in the sense that
$\bra{\alpha} \bra{s} \mathcal{A}^k \ket{\beta} \ket{s'} \equiv [\mathcal{A}^k]^{\alpha s}_{\beta s'}$.
$\alpha$ and $\beta$ are the virtual indices that range from $0$ to $D-1$ (with $D$ being the virtual dimension of the state).
One the other hand, $s$ and $s'$ are the symmetry indices that transfer the information about the number
of down-spins from left to right. The local constraint that guarantees this information transfer is
$s' = s + k$. This constraint determines the blocks $[\mathcal{A}^k]^{- s}_{- s'}$ that are non-zero
and allows a sparse storage of the state.
The left boundary-state $\bra{0}$ and the right boundary-state $\ket{m}$ fix the
total number of down-spins of the MPS to~$m$.
The optimization problem~\ref{eqn:optproblem} at step~$m$ then
consists in approximating the state $\ket{\Psi_m}$ with $m$ down-spins by a state $\ket{\tilde{\Psi}_m}$
that also has $m$ down-spins. This leads to a gain of a factor of~$m$ in time and memory.

\begin{figure}[t]
    \begin{center}
        \includegraphics[width=0.48\textwidth]{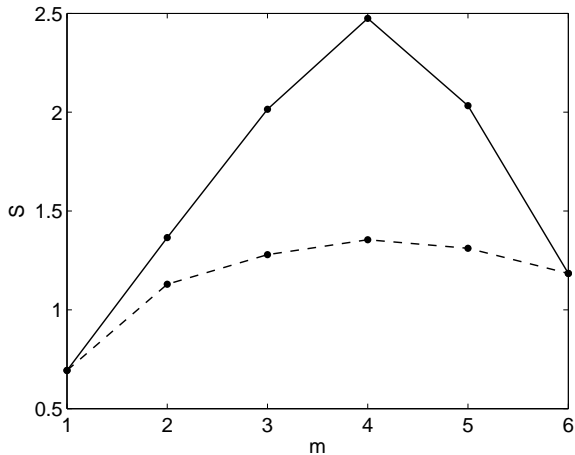}
     \end{center}
    \caption{
	Half-chain entropy as a function of the evolution step for
	the ground state of the $N=12$ Heisenberg model with periodic boundary conditions.
	The lower line is obtained with the orderings $(435261)$, $(435216)$, $(432561)$, $(432516)$, $(345261)$, $(345216)$, $(342561)$, $(342516)$;
	the upper line is obtained with the orderings $(531624)$, $(426153)$, $(351624)$, $(246153)$.
       }
    \label{fig:allperms}
\end{figure}

Furthermore, there is a (mathematical) degree of freedom that
can be used to improve the approximation.
This degree of freedom is due to the commutativity-property of $B(\mu)$:
since $[B(\mu),B(\nu)]=0$ for all $\mu$ and $\nu$, the ordering of the
$B(\mu_j)'s$ in the Ansatz~(\ref{eqn:betheansatz}) is completely arbitrary.
This is relevant insofar as the entanglement-properties of the
intermediate states~$\ket{\Psi_m}$ ($1 < m  < M$) are concerned.
The intermediate states are a priori no physical ground states, i.e. 
there is no reason for them to lie in the set of MPS with low
bond-dimension. However, as we see numerically, there is always an ordering
such that the intermediate states contain as little entanglement as possible.
This ordering we then use for calculating the approximation.
That the ordering has a formidable effect can be gathered from Fig.~\ref{fig:allperms}
for the example of the ground state of the $12$-site Heisenberg antiferromagnet with periodic boundary conditions.
Here, the half-chain entropy of~$\ket{\Psi_m}$ is plotted
as a function of~$m$ for the best and the worst ordering.
As it can be seen, the entropy is highest at the intermediate steps and
decreases at $m \to M$ when the state becomes a physical ground state.

\begin{figure}[t]
    \begin{center}
        \includegraphics[width=0.48\textwidth]{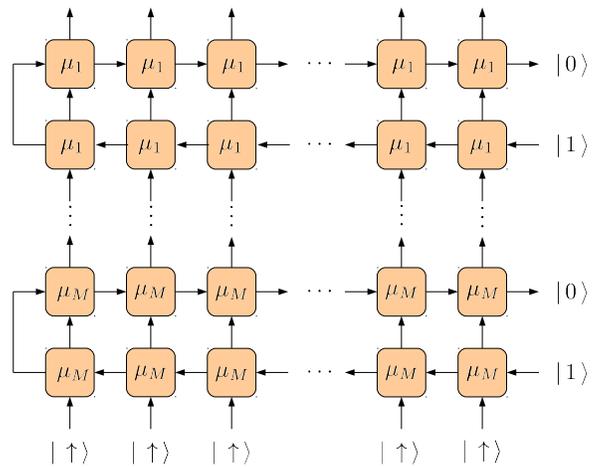}
     \end{center}
    \caption{
	Tensor network constituting the Bethe eigenstate of the
	Heisenberg model or XXZ model with open boundary conditions.
       }
    \label{fig:betheansatzopenbc}
\end{figure}

Up to now, we have considered the Bethe Ansatz for the XXZ model with periodic boundary conditions
(and models in this class). 
In case of open boundary conditions, the Bethe Ansatz has the same form as in~$(\ref{eqn:betheansatz})$,
merely the creation Operators are not single MPOs, but products of two MPOs~\cite{cherednik84,sklyanin88,murg12}:
\begin{displaymath}
\mathcal{B}(\mu) = \sum_{s=0}^1 \bar{B}_s(\mu) B_{1-s}(\mu)
\end{displaymath}
$B_{1-s}(\mu)$ has the property to create $1-s$ down-spins, whereas $\bar{B}_{s}(\mu)$ creates $s$ down-spins ($s \in \{0,1\}$),
such that $\mathcal{B}(\mu)$ is a creation operator for exactly one down-spin, as before.
The tensor-network representation for the Bethe-state with open boundary conditions is shown in Fig.~\ref{fig:betheansatzopenbc}.
It contains twice as many rows as the tensor-network for periodic boundary conditions,
which makes the contraction more challenging, in principle.
However, as we see numerically, after a multiplication with a MPO-pair~$\mathcal{B}(\mu)$,
the Schmidt-rank of the state increases only by a factor of~$2$ - not $4$, as expected.
Thus, the numerical effort to contract the tensor-network is similar.


\begin{figure}[t]
    \begin{center}
	\includegraphics[width=0.48\textwidth]{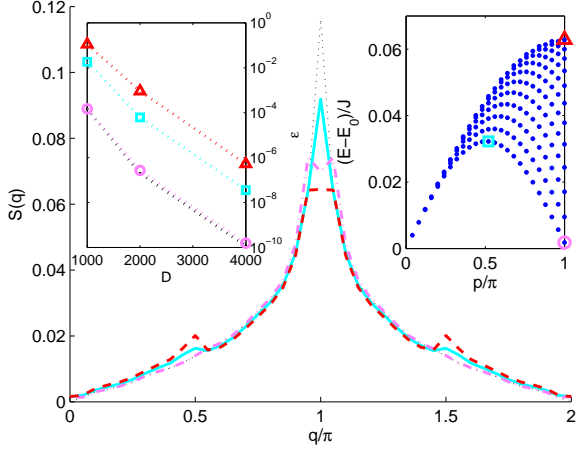}
     \end{center}
    \caption{
	Structure factor for the ground state (dotted line) and three selected two-spinon 
	excited states of the periodic $N=50$-Heisenberg chain
	as a function of the wave-vector~$q$ obtained from approximated Bethe-MPS with $D=1000$.
	The three states are marked in the inset a the rhs which
	shows all energies and momenta 
	of the two-spinon excited states with total spin $S=1$ and total $z$-spin $S_z=1$
	The inset at the lhs depicts the relative error
	in the energy as a function of~$D$ for the three selected two-spinon states.
       }
    \label{fig:sq3statesN50}
\end{figure}

\begin{figure}[t]
    \begin{center}
        \includegraphics[width=0.48\textwidth]{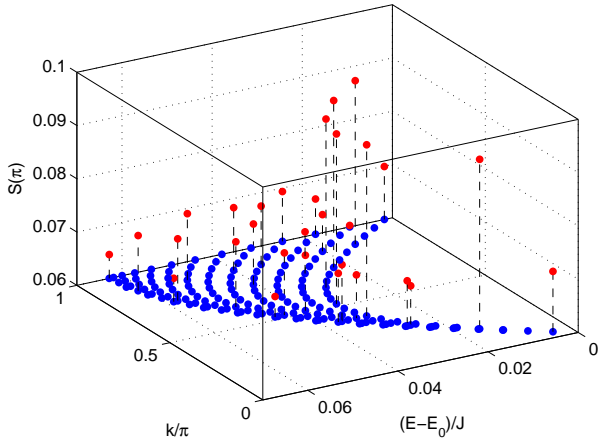}
     \end{center}
    \caption{
	Structure factor~$S(\pi)$ for selected two-spinon excited states of the
	$N=50$ Heisenberg antiferromagnet as a function of the momentum and the excitation energy.
	The used cutoff is~$D=1000$.
       }
    \label{fig:sqN50}
\end{figure}

Using the previously described method, we have obtained results for the
Heisenberg model with periodic boundary conditions and the XXZ-model
with open boundary conditions.

In case of the Heisenberg model, we have investigated the two-spinon
excited states with total spin $S=1$ and total $z$-spin $S_z=1$.
The energies of these states obtained by the Bethe-Ansatz
as a function of the momentum for $N=50$ spins are plotted in
the right inset in Fig.~\ref{fig:sq3statesN50}.
With respect to these states, we have calculated the 
correlation functions $\expect{\sigma_z^{r} \sigma_z^{s}}$ and
the corresponding structure factor
\begin{displaymath}
S_z(q) = \frac{1}{N^2} \sum_{r,s} e^{i q (r-s)} \expect{\sigma_z^{r} \sigma_z^{s}}.
\end{displaymath}
The structure factor $S_z(q)$ as a function of $q$ for the three selected excited states
(that are marked in the right inset)
can be gathered from Fig.~\ref{fig:sq3statesN50}.
In order to judge the accuracy of the approximated Bethe-MPS
obtained with our method,
we have compared the the expectation values of the
Hamiltonian with respect to these MPS
to the energies obtained from the Bethe-Ansatz.
For the case $N=50$, the so obtained relative error plotted as a function of~$D$
for the selected excited states can be gathered from 
the left inset in Fig.~\ref{fig:sq3statesN50}.
As can be seen, the error decreases with increasing~$D$, but also
with decreasing excitation energy.
In Fig.~\ref{fig:sqN50}, the structure factor at the point $q=\pi$, i.e. the
squared staggered magnetization, is plotted as a function of the
excitation energy and the momentum.
Evidently, the excited states of the lowest branch show the highest staggered magnetization.

\begin{figure}[t]
    \begin{center}
        \includegraphics[width=0.48\textwidth]{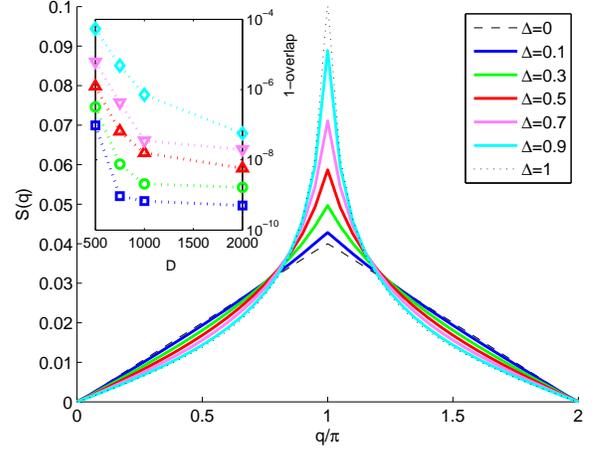}
     \end{center}
    \caption{
    	Structure factor~$S(q)$ of the ground states of the
    	$N=50$ XXZ-model with open boundary conditions for selected values of $\Delta$
	obtained from approximated Bethe-MPS with $D=1000$.
	The dashed black line marks the free fermion case $\Delta=0$ and the dotted black line
	indicates the Heisenberg limit $\Delta=1$.
	The inset shows the overlap with MPS of virtual dimension~$D=400$ obtained from DMRG
	calculations.
       }
    \label{fig:sq2xxz}
\end{figure}

In case of the XXZ-model with open boundary conditions, we have studied the correlations of the ground state.
To get an impression of the quality of the obtained result, we computed the overlap with MPS states
obtained from DMRG calculations. 
For the case of~$50$ spins, the overlap
as a function of~$D$ can be gathered from the inset of Fig.~\ref{fig:sq2xxz} for
different values of~$\Delta$.
The structure factor $S(q)$ as a function of the wave-vector~$q$ for different values of $\Delta$ 
evaluated with respect to the approximated Bethe-MPS with $D=1000$
is plotted in the main part of Fig.~\ref{fig:sq2xxz}.
The structure factor obtained for this~$D$ only deviates marginally from the
structure factor obtained from the DMRG calculation.


Summing up, we have presented a method for approximative calculation of correlation functions with
Bethe-eigenstates. For this, we make use of the fact that a Bethe eigenstate is a product
of MPOs applied to a MPS. We systematically reduce the virtual dimension after each
multiplication and obtain an MPS with small virtual dimension that can be used for
the calculation of any expectation value.
We have shown the effective operation of this method by applying it to the Heisenberg antiferromagnet
with periodic boundary conditions and the XXZ model with open boundary conditions.
We have obtained results for the structure factor of ground- and excited states and
compared our ground state-results to DMRG calculations.


\acknowledgments{
V.~M.\ and F.~V.\ acknowledge support from the SFB projects
FoQuS and ViCoM, the European projects Quevadis, and the ERC grant
Querg.
V.~K achnowledges support from the NSF grant Grant DMS-0905744.
}



%


\clearpage
\onecolumngrid
\appendix

\section{Supplementary Material}

In this supplementary material, we describe the numerical method used in our article in detail.

\section{Numerical State Approximation}

The main building block of the algorithm is the approximation of MPS~$\ket{\Psi_m}$ with a fixed number of~$m$ down-spins
and virtual dimension~$D$ by a MPS~$\ket{\tilde{\Psi}_m}$ that also has~$m$ down-spins
and virtual dimension~$\tilde{D}<D$ in a way that the distance~(\ref{eqn:optproblem}) between 
the two states is minimal. The state~$\ket{\Psi_m}$ reads
\begin{displaymath}
\ket{\Psi_m} = \sum_{k_1 \cdots k_N} \bra{0} \bra{0} \mathcal{A}^{k_1} \cdots \mathcal{A}^{k_N} \ket{0} \ket{m} \ket{k_1,\ldots,k_N}
\end{displaymath}
with $\bra{\alpha} \bra{s} \mathcal{A}^k \ket{\beta} \ket{s'} \equiv [\mathcal{A}^k]^{\alpha s}_{\beta s'}$
and $s' = s + k$.
The MPS $\ket{\tilde{\Psi}_m}$ has the same structure, but may be inhomogeneous, i.e. the matrices may be site-dependent:
\begin{displaymath}
\ket{\tilde{\Psi}_m} = \sum_{k_1 \cdots k_N} \bra{0} \bra{0} \tilde{\mathcal{A}}^{k_1}_1 \cdots \tilde{\mathcal{A}}^{k_N}_N \ket{0} \ket{m} \ket{k_1,\ldots,k_N}
\end{displaymath}

We obtain the starting point for the optimization by setting $\ket{\tilde{\Psi}_m}$ equal to $\ket{\Psi_m}$
and performing Schmidt-decompositions successively for bonds $(1,2)$ to $(N-1,N)$, where we
keep at each bond only the~$\tilde{D}$ largest Schmidt-coefficients.
Due to the conservation of the number of down-spins,
the Schmidt-decomposition with respect to one bond~$(j,j+1)$ can be written as
\begin{displaymath}
\ket{\tilde{\Psi}_m} =
\sum_{s=0}^m \sum_{\gamma=0}^{\Gamma_s} \sigma^{s}_{\gamma} \ket{\Phi^s_{\gamma}} \otimes \ket{\bar{\Phi}^{m-s}_{\gamma}}.
\end{displaymath}
Thereby, the states $\{ \ket{\Phi^s_{\gamma}} | \gamma=1,\ldots,\Gamma_s\}$ are states acting on 
the left block (sites $1,\ldots,j$) with a 
fixed number of~$s$ down-spins. In addition, these states satisfy the orthonormality constraint
$\scal{\Phi^s_{\gamma}}{\Phi^s_{\gamma'}} = \delta_{\gamma \gamma'}$.
In an analogous manner, $\{ \ket{\bar{\Phi}^{m-s}_{\gamma}} | \gamma=1,\ldots,\Gamma_s \}$ are orthonormal states acting 
on the right block (sites $j+1,\ldots,N$) with 
a fixed number of~$m-s$ down-spins.
The singular values corresponding to the partitioning of $s$ down-spins in the left block and $m-s$ down-spins in the right block are
$\{ \sigma^{s}_{\gamma} | \gamma=1,\ldots,\Gamma_s \}$.

The Schmidt-decomposition can be obtained by transforming the MPS into the form
\begin{displaymath}
\ket{\tilde{\Psi}_m} = \sum_{k_1 \cdots k_N} \bra{0} \bra{0} 
\cdot \cdot \cdot
\tilde{\mathcal{A}}^{k_j}_j \Sigma \tilde{\mathcal{A}}^{k_{j+1}}_{j+1}
\cdot \cdot \cdot
\ket{0} \ket{m} \ket{k_1,\ldots,k_N},
\end{displaymath}
with $\tilde{\mathcal{A}}^k_i$ fulfilling the local constraints
$\sum_k (\tilde{\mathcal{A}}^k_i)^{\dagger} \tilde{\mathcal{A}}^k_i = \mathbb{1}$ for $i=1,\ldots,j$
and
$\sum_k \tilde{\mathcal{A}}^k_i (\tilde{\mathcal{A}}^k_i)^{\dagger} = \mathbb{1}$ for $i=j+1,\ldots,N$.
The local constraints guarantee the orthonormality of the states in the left and right block.
$\Sigma$ is a diagonal matrix containing the Schmidt-coefficients.
The transformation into this form is always possible due to the gauge invariance of MPS~\cite{verstraeteciracmurg08}.

For $i \leq j$, the way to meet the local constraints is
by QR-decompositions of the matrices 
$\hat{\mathcal{A}}^{s'}_i$ defined as
$[ \hat{\mathcal{A}}^{s'}_i ]^{\alpha s k}_{\beta} \equiv [ \tilde{\mathcal{A}}^k_i ]^{\alpha s}_{\beta s'}$, successively for $i=1,\ldots,j$:
$\hat{\mathcal{A}}^{s'}_i = Q^{s'}_i R^{s'}_i$.
The matrix $Q^{s'}_i$ is an appropriate replacement for $\hat{\mathcal{A}}^{s'}_i$, because the
orthogonality of $Q^{s'}_i$ makes the local constraint satisfied.
To keep the MPS invariant,
the block-diagonal matrix~$R_i$ defined as $[R_i]^{\alpha s}_{\beta s'} \equiv [R_i^{s'}]^{\alpha}_{\beta} \delta_{s s'}$
must be used to update $\tilde{\mathcal{A}}^k_{i+1}$ to $R_i \tilde{\mathcal{A}}^k_{i+1}$.
Only at the last step $i=j$, the update must be omitted.
For $i>j$, LQ-decompositions are performed in an analogous manner:
successively for $i=N,\ldots,j+1$, the matrices 
$[ \hat{\mathcal{A}}^{s}_i ]^{\alpha}_{\beta s' k} \equiv [ \tilde{\mathcal{A}}^k_i ]^{\alpha s}_{\beta s'}$
are decomposed as $L^s_i Q^s_i$.
The matrices $Q^s_i$ are then used to replace $\hat{\mathcal{A}}^{s}_i$ and
$L^s_i$ update $\tilde{\mathcal{A}}^{k}_{i-1}$ to
$\tilde{\mathcal{A}}^{k}_{i-1} L_i$ (with $[L_i]^{\alpha s}_{\beta s'} \equiv [L_i^{s}]^{\alpha}_{\beta} \delta_{s s'}$).
As before, the update must be omitted at the last step $i=j+1$.

The Schmidt-decomposition is now obtained by
performing a singular value decomposition of the product of the ``left-over'' update-matrices $R^s_{j}$ and $L^s_{j+1}$:
$R^s_{j} L^s_{j+1} = U^s \Sigma^s V^s$.
$U^s$ and $V^s$ are unitary, as well as their block-diagonal extensions $U$ and $V$ defined as
$[U]^{\alpha s}_{\beta s'} \equiv [U^s]^{\alpha}_{\beta} \delta_{s s'}$ and
$[V]^{\alpha s}_{\beta s'} \equiv [U^s]^{\alpha}_{\beta} \delta_{s s'}$.
Because of their unitarity, they can be used to update $\tilde{\mathcal{A}}^k_j$ to $\tilde{\mathcal{A}}^k_j U$
and $\tilde{\mathcal{A}}^k_{j+1}$ to $V \tilde{\mathcal{A}}^k_{j+1}$ without spoiling the orthonormality constraint.
The matrix $\Sigma^s$ contains the Schmidt-coefficients $\{ \sigma^s_{\gamma} | \gamma=1,\ldots,\Gamma_s \}$
related to the partitioning of $s$ down-spins in the left block and $m-s$ down-spins in the right block.
With the definition of $\Sigma$ as $[\Sigma]^{\alpha s}_{\beta s'} \equiv [\Sigma^s]^{\alpha}_{\beta} \delta_{s s'}$,
we have obtained the desired form of the MPS.

A sensible way to reduce the dimension of bond $(j,j+1)$ is keeping only the largest~$\tilde{D}$ Schmidt-coefficients
and setting all others to zero. Thus, by defining a projector $P$, such that $P \Sigma P^{\dagger}$ is the $\tilde{D} \times \tilde{D}$-matrix
containing the~$\tilde{D}$ largest Schmidt-coefficients and updating $\tilde{\mathcal{A}}^k_{j}$ to $\tilde{\mathcal{A}}^k_{j} P^{\dagger}$
and $\tilde{\mathcal{A}}^k_{j+1}$ to $P \Sigma \tilde{\mathcal{A}}^k_{j+1}$, we obtain
the favoured MPS with reduced bond-dimension.

Performing the Schmidt-decompositions with successive projections for all bonds $(1,2)$ to $(N-1,N)$
gives a MPS that is a fairly good starting point for the optimization problem~($\ref{eqn:optproblem}$).
A further improvement is possible by optimizing the quantity~$K$ locally, i.e.
by optimizing~$K$ with respect to the matrices~$\{\tilde{\mathcal{A}}^k_j | k=0,1\}$ at one site~$j$,
and keeping all other matrices constant. This has already been discussed extensively in~[\onlinecite{murgverstraete05},\onlinecite{verstraeteciracmurg08}]. The main idea is that~$K$ is a quadratic function of $\tilde{\mathcal{A}}^k_j$, such that
it can be written as
\begin{displaymath}
K = \mathrm{const} + \sum_{k k'} (\tilde{\mathcal{A}}^k_j)^{\dagger} \mathcal{N}^{k k'}_j \tilde{\mathcal{A}}^{k'}_j
+ \sum_k (\tilde{\mathcal{A}}^k_j)^{\dagger} w^k_j.
\end{displaymath}
The minimum with respect to $\tilde{\mathcal{A}}^k_j$ is achieved for those values of $\tilde{\mathcal{A}}^k_j$ solving
the system of linear equations 
\begin{displaymath}
\sum_{k'} \mathcal{N}^{k k'}_j \tilde{\mathcal{A}}^{k'}_j = w^k_j.
\end{displaymath}
The matrix $\mathcal{N}^{k k'}_j$ is a function of $\tilde{\mathcal{A}}^{k}_i$, $i \neq j$, and
it is equal to the identity if the constraints
$\sum_k (\tilde{\mathcal{A}}^k_i)^{\dagger} \tilde{\mathcal{A}}^k_i = \mathbb{1}$ for $i=1,\ldots,j-1$
and
$\sum_k \tilde{\mathcal{A}}^k_i (\tilde{\mathcal{A}}^k_i)^{\dagger} = \mathbb{1}$ for $i=j+1,\ldots,N$.
are fulfilled.
These constraints, however, can always be imposed, as argumented earlier.
By performing the local minimization for~$j$ sweeping between~$1$ and~$N$ until convergence of~$K$,
the global minimum of~$K$ is (usually) obtained and we have found the optimal approximation
$\ket{\tilde{\Psi}_m}$ with maximal bond-dimension~$\tilde{D}$ to the state $\ket{\Psi_m}$.



\end{document}